\documentclass[]{aa}  
\usepackage{txfonts}
\usepackage[colorlinks=true,citecolor=blue,urlcolor=blue,breaklinks=true]{hyperref}
\usepackage{graphicx,epstopdf}
\DeclareGraphicsExtensions{.pdf}
\usepackage{txfonts}
\usepackage{lipsum}
\usepackage{placeins}   
\usepackage{subcaption}         
\usepackage{lscape}             

\begin{document}

   \title{The Accretion Properties and Jet Mechanisms for the Low-Excitation Radio Galaxies}
\titlerunning{Accretion and Jets for LERGs}
\authorrunning{Ye et al.}

   \author{Xu-Hong Ye\inst{1,2,3}
        \and Ranieri D. Baldi\inst{4}
        \and Yong-Yun Chen\inst{5} 
        \and Denis Bastieri \inst{1,2,3}
        \and Jun-Hui Fan \inst{3,6}
        }

   \institute{Dipartimento di Fisica e Astronomia “G. Galilei,” Università di Padova, Via F. Marzolo, 8, I-35131 Padova, Italy\\
             \email{denis.bastieri@unipd.it}
            \and Istituto Nazionale di Fisica Nucleare, Sezione di Padova, I-35131 Padova, Italy
            \and Centre for Astrophysics, Guangzhou University, Guangzhou 510006, People's Republic of China \\
            \email{fjh@gzhu.edu.cn}
            \and INAF - Istituto di Radioastronomia, via Gobetti 101, 40129 Bologna, Italy
            \and College of Physics and Electronic Engineering, Qujing Normal University, Qujing 655011, People's Republic of China 
            \and Astronomy Science and Technology Research Laboratory of Department of Education of Guangdong Province, Guangzhou 510006, People's Republic of China
            }

   \date{Received September 30, 20XX}

 
  \abstract
   {Radio galaxies (RGs) are a subclass of active galactic nuclei, which are suggested to be the parent populations of blazars.  According to the accretion-ejection paragram, RGs can be classified into low-excitation or high-excitation radio galaxies (LERGs or HERGs).}
   {In this paper, we compiled a distance-limited ($z<0.15$) sample of 431 LERGs (Fanaroff-Riley, or FR, type 0, I, and II RGs) to discuss their jet formation mechanism with the ADAF (advection-dominated accretion flow) scenario, and compare their accretion properties with Fermi BL Lacertae objects (BL Lacs).} 
   {We explored different jet mechanisms (Blandford-Znajek [BZ] model and a mixture of the BZ and Blandford-Payne, hybrid, model) within the framework of ADAF-type disc around a Kerr black hole for both LERGs and Fermi BL Lacs.}
   {Based on standard assumptions on the accretion-ejection coupling in RGs, the maximum kinetic jet and accretion power for FR 0s, FR Is, FR IIs can be,  explained by an ADAF with the pure BZ mechanism or hybrid jet mechanism. In addition, for one third of the FR IIs, to account for their higher kinetic jet power than what is simply expected by the hybrid jet mechanism, the magnetic field could play an important role as in the form of magnetization-driven outflows or stronger magnetic structures, as observed in some BL Lacs with high jet powers.}
   {Similarities between BL Lacs and LERGs (e.g., accretion-ejection and clustering properties) suggest that high synchrotron peaked BL Lacs could be the beamed counterparts of FR 0s, and a potential general unification between LERGs and BL Lacs populations is discussed. However, a complete sample of BL Lacs is needed to robustly compare the jet and accretion properties with those of LERGs in the future.}

  {}


\maketitle
   
\section{Introduction}
\label{intro}
Radio galaxies (RGs), a subclass of radio-loud active galactic nuclei (AGNs), are classified into Fanaroff-Riley type I (FR Is) or type II (FR IIs) RGs based on morphologies and radio powers \citep{fan74}. However, the FR dichotomy for RGs is not strictly separated by a typical value of the extended radio power ($L_{\rm{ext}}$) \citep{owe94,zirbel95}; FR Is have lower radio powers ($L_{\rm{ext}}<10^{24.5}$ W/Hz), while FR IIs have higher radio powers ($L_{\rm{ext}}>10^{26}$ W/Hz).  An intermediary region ($10^{24.5}<L_{\rm{ext}}<10^{26}$ W/Hz) is occupied by both FR Is and FR IIs.  Besides the radio sources extended up to Mpc (megaparsec) scale, the compact radio sources have also been studied for decades (e.g. \citealt{condon78,kellermann81,falcke04}). The population of compact radio sources dominates the local Universe (e.g. \citealt{odea21}). A large fraction of these compact RGs lack of kiloparsec (kpc) extended structure, and are not interpreted as young RGs: these have been named as Fanaroff-Riley type 0 RGs (FR 0s) \citep{bal15, bal23}. The FR 0s share similar nuclear and host properties to FR Is, but show total luminosities lower than classical FR Is and FR IIs by a factor 100-1000 \citep{bal15}, not fitting into the traditional FR I-II classification.

\cite{bes12} suggested a fundamental classification for RGs using the accretion-ejection paradigm, where different structures and different environments surrounding the black hole lead to different accretion mechanisms. Radio galaxies with a 'standard' geometrically thin and optically thick disc \citep{shakura73}, showing a higher accretion rate ($\dot{m}$, typically 1-10$\%$ of their Eddington rate) on to the supermassive black holes (BHs), are classified as high-excitation radio galaxies (HERGs); in contrast, RGs with low accretion rates, dominated by a radiatively inefficient disc, are named as low-excitation radio galaxies (LERGs)  \citep{hec14}.  At the centre of LERGs, the thin accretion disc is replaced by an advection-dominated accretion flow (ADAF) \citep{nar95, hec14}. Most FR 0s and FR Is are LERGs,  whereas FR IIs exhibit varying accretion properties \citep{bal08,hec14,cap17a, tor18}. Some FR IIs have high accretion rates, known as FR II HERGs, while others display low accretion disc properties, called FR II LERGs \citep{har09,mingo22}. In general, HERGs are homogeneous with FR IIs, while LERGs are heterogeneous with both FR Is and FR IIs \citep{but10}.
  
Blazars are a subclass of radio-loud AGNs showing extreme observation properties with the relativistic jets closely aligned to the line of sight: rapid variabilities, high polarizations, high luminosities, superluminal motions, radio core-dominance morphologies, or highly energetic radiations \citep{urr95, xia19, pei20, fan16,fan21, hom21, aje22, lio22, yuan22, liang23, zhang23}. There are two subclasses of blazars:  BL Lacertae objects (BL Lacs) with weak equivalent emission-line widths (EW $<5\mathring{A}$) and flat-spectrum radio quasars (FSRQs) with significant equivalent emission-line widths (EW $>5\mathring{A}$) \citep{sti91}. The classification scheme for (misaligned) RGs and (aligned) blazars are still inconsistent. Radio galaxies are typically classified based on their radio morphology and radio power \citep{fan74}, whereas classes of blazars are classified by the strengths of the emission lines \citep{sti91}. The classical picture was that FR Is and FR IIs are the parental populations of BL Lacs and FSRQs, respectively. 

Radio-loud AGNs are characterised by a non-thermal synchrotron spectrum, generated from accretion-ejection physical processes from the central engine, which has a typical luminosity peak at a specific radio-band peak frequency where the emission moves from an optically-thick regime to an optically-thin one. Based on the synchrotron peak of the spectral energy distribution (SED), blazars are classified into low synchrotron peaked blazars ($\log\nu_{\rm{peak}}<13.7$ Hz), intermediate synchrotron peaked blazars ($13.7<\log\nu_{\rm{peak}}<14.9$ Hz), and high synchrotron peaked blazars ($\log\nu_{\rm{peak}}>14.9$ Hz) \citep{yang22}. Most FSRQs are low synchrotron peaked sources, while BL Lacs can be classified into low, intermediate, and high synchrotron peaked BL Lacs, as noted LBLs, IBLs, and HBLs \citep{aje22}. In the plane of the synchrotron peak frequency-peak luminosity ($\nu_{\rm{peak}}-L_{\rm{peak}}$), an anti-correlation between them for blazars (so-called blazar sequence) was first discussed in \cite{fos98}. The blazar sequence is explained by the Compton cooling effect \citep{ghi98, pra22}, selection effects \citep{gio12}, or beaming effects \citep{nie08, yangwx22}. However, \citet{kee21} found little evidence of the anti-correlation for the blazar sequence, but they found a dichotomy: stronger jets and more efficient discs in HERGs, FSRQs and most LBLs than LERGs and blazars with $\nu_{\rm{peak}}>10^{15}$ Hz.

For radio-loud AGNs, there are two main scenarios explaining the jet formation within the accretion disc and/or spinning BH \citep{bz77, bp82}. If jets extract the rotational energy of black holes, the jet formation mechanism follows the Blandford-Znajek (BZ) mechanism \citep{bz77}. If the jets extract the rotational energy from the accretion disc, it follows the Blandford-Payne (BP) mechanism \citep{bp82}. The jet power difference between the pure BZ and BP models is primarily determined by the BH spin and the self-similar index, which describes how the poloidal magnetic field varies with the cylindrical radius of the jet. Nevertheless, at a given magnetic field, for a rapid BH spin (e.g. $j > 0.9$),  the jet power of BZ model is larger than that of the BP model for any magnetic field configuration \citep{li08}. 
Both of BZ and BP models contribute to the physics of the disc, leading to an expected relationship between the disc luminosity and jet power \citep{ghi11,sba12}.  If the rotational energy of the jet is from both accretion discs and black holes, the jet formation mechanism could be a mixture (hybrid model) of BP and BZ mechanism  \citep{mei01,nem07}. 
In this paper, we compiled a nearby sample ($z<0.15$) of 431 LERGs (FR 0s, FR Is and FR IIs) to discuss their jet mechanisms in the case of ADAFs and compare their accretion properties with those of the Fermi BL Lacs. A $\Lambda$-CMD cosmology with $\Omega_{\Lambda}\sim0.7$, $\Omega_{\rm{M}}\sim0.3$, and $H_0$ = 70 km s$^{-1}$ Mpc$^{-1}$ is applied throughout the paper.

\section{Sample} 
\cite{bes12} compiled a sample of 18,286 radio-loud AGNs by cross-checking several datasets from the Sloan Digital Sky Survey (SDSS), the National Radio Astronomy Observatory (NRAO) Very Large Array (VLA) Sky Survey (NVSS), and the Faint Images of the Radio Sky at Twenty centimetres (FIRST) survey, and classified the sources into LERGs or HERGs based on their emission-line properties, applying multiple classification criteria (e.g. \citealt{kew06,but10,cid10}).
Within this sample, \citet{cap17a, cap17b} searched for low-power FR Is (FRICAT) and FR IIs (FRIICAT) by using the morphological classifications of the FIRST radio images. They visually inspected the FIRST images of each radio source with a flux density larger than 5 mJy, and limited the sample with a redshift of $z<0.15$. All sources with radio emissions extending at least 30 kpc are preserved. The 30 kpc radius corresponds to $11''.4$ for the distant sources, and ensures that all the FRICAT/FRIICAT sources are well resolved with the $5''$ resolution of the FIRST images, enabling a detailed exploration of their morphologies. FRICAT sources have one- or two-sided jets that fade in brightness along their length without brightening at the ends, while FRIICAT sources show edge-brightened morphology peaking at least 30 kpc from the optical host centre.
They obtained a sample of 327 RGs, including 219 FR I LERGs and 108 FR II LERGs, which we considered in this work.

\cite{bal18} followed the same procedures outlined by \citet{cap17a, cap17b}, selecting the sources with $z<0.05$,  a minimum flux density of 5 mJy, and a maximum offset of $2''$ from the optical centre. They visually inspected the FIRST images, and eliminated the sources with clearly extended radio emissions to select only the compact radio sources. Sources with an observed major axis smaller than $6''.7$ were preserved, corresponding to a size of 5 kpc at $z=0.05$. Finally, \cite{bal18} compiled a sample of 108 FR 0 LERGs, in which 4 FR 0 LERGs are discarded based on the images of high-resolution radio observations by \citet{bal19}.

In summary,  we compiled a total of 431 LERGs, of which 104 FR 0s, 219 FR Is and 108 FR IIs. This sample consists of low-power radio-loud AGNs ($>$ 5 mJy at 1.4 GHz), weaker than the classical 3CR (Third Cambridge Revised Catalogue) RGs (e.g. \citealt{ben62,spinrad85}, selected at $>$9 Jy at 178 MHz), and their hosts have $r$-band magnitudes in the range of $15.5 < r < 13$, covering a redshift completeness of the $\sim90\%$ of SDSS optical main galaxies sample \citep{strauss02}. The 431 LERGs with available redshifts, 1.4 GHz NVSS radio luminosities, emission-line luminosities, jet powers, and BH masses are presented in Table \ref{tab:1}.

\begin{table*}[hbtp]
	\centering
	\caption{The parameters for nearby 431 low-excitation radio galaxies and 75 Fermi BL Lacertae objects.}
	\label{tab:1}
	\begin{tabular}{ccccccccc} 
 \hline
 
 Name&class&$z$ &$L_{\rm{1.4}}$ &$\log L_{\rm{kin}}$&$\log L_{\rm{line}}$& $\log M_{\rm{BH}}$& $\lambda$\\
 &&&erg/s&erg/s&erg/s&&\\
 (1)&(2)&(3)&(4)&(5)&(6)&(7)&(8)\\
 \hline
 J0003.2+2207&HBL&0.1&39.42&43.28&41.72&8.10&-4.52\\
  J002900.98-011341.7
&FR I&0.083&40.80&44.21&42.61&8.96&-4.52\\
J003930.52-103218.6&FR I&0.129&40.13&43.75&41.10&8.4&-5.44\\
J010852.48-003919.4&FR 0&0.045&38.83&42.87&41.93&8.32&-4.51\\
J001247.57+004715.8&FR II&0.148&40.65&44.11&42.01&8.61&-4.73\\

 $\ldots$&$\ldots$&$\ldots$&$\ldots$&$\ldots$&$\ldots$&$\ldots$&$\ldots$\\
\hline

	\end{tabular}
\small

Notes: Col. (1) is the source name; Col. (2) the classification. BL Lacs are classified into low, intermediate, and high synchrotron peaked BL Lacs, as noted LBLs, IBLs, and HBLs; low-excitation radio galaxies are classified into Fanaroff-Riley type 0 (FR 0), I (FR I), and II (FR II) radio galaxies; Col. (3) the redshift; Col. (4) the 1.4 GHz luminosity (in erg/s). For BL Lacs, the intrinsic 1.4 GHz luminosity is calculated by assuming a Doppler factor $\delta \sim 10$ within the continuous jet model; Col. (5) the kinetic jet power (in erg/s) derived from 1.4 GHz radio luminosities. Col. (6) the total emission-line luminosity (in erg/s); Col. (7) The BH masses in the unit of solar mass derived from the $M-\sigma_\ast$ relation; Col. (8) the line accretion rate $\lambda=L_{\rm{line}}/L_{\rm{Edd}}$. 

(A portion of the table is presented here to offer guidance on the format and content of Table 1.)
\end{table*}

\subsection{Redshift}
\begin{figure}
    \centering
    \includegraphics[width=0.9\linewidth]{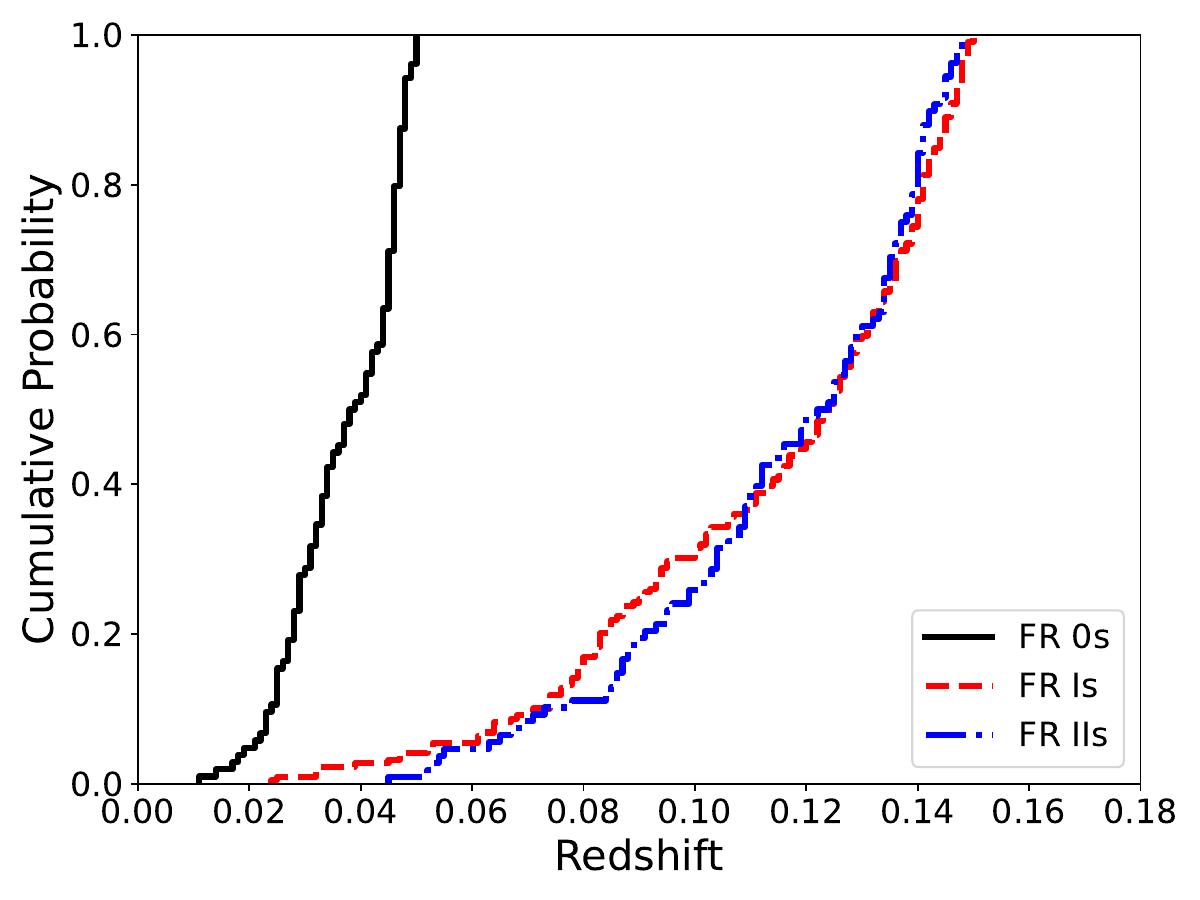}
    \caption{The cumulative distributions of the redshifts for 431 low-excitation radio galaxies. The black solid line is for FR 0s, the red dashed line is for FR Is, and the blue dash-dotted line is for FR IIs. The average redshift for FR 0s is $\langle z_0\rangle=0.037\pm0.001$, whereas that of FR Is is $\langle z_{\rm{I}}\rangle=0.114\pm0.002$, and that of FR IIs is $\langle z_{\rm{II}}\rangle=0.115\pm0.003$. }
    \label{fig:z}
\end{figure}

The redshift range for FR 0s is from $z = 0.011$ to $z = 0.05$, with an average redshift of $\langle z_0\rangle= 0.037 \pm 0.001$. In comparison, FR Is and FR IIs have higher redshift ranges: FR Is range from $z = 0.024$ to $z = 0.15$ with an average redshift of $\langle z_{\rm{I}}\rangle= 0.114 \pm 0.002$, and FR IIs range from $z = 0.045$ to $z = 0.148$ with an average redshift of $\langle z_{\rm{II}}\rangle= 0.115 \pm 0.003$. The redshifts for the 431 LERGs are presented in Fig. \ref{fig:z} and Table \ref{tab:1}.

The Kolmogorov-Smirnov (K-S) test results suggest that FR Is and FR IIs have similar redshift distributions (K-S test $p = 0.09$). However, both of these distributions are significantly larger than the redshift distribution of FR 0s (both $p<0.05$).

\subsection{Kinetic jet power}
For large-scale radio sources, a portion of the total jet power is either converted into kinetic energy or dissipated due to the work done by the expanding lobes \citep{sch74, wil99}. The jets push the relativistic plasma to interact with the surrounding thermal gas, appearing as bubbles or cavities in X-ray observations (e.g. \citealt{boh93}). \citet{wil99} estimated the kinetic (mechanical) jet power for FR IIs from the 151 MHz radio luminosity, assuming that the lobes store the minimum energy required to produce the observed synchrotron emission over an age of $10^7$ years.  \citet{cao04} suggested that this estimation is also applicable for FR Is, with an acceptable accuracy of an order of magnitude compared to other methods (e.g. M87; \citealt{bicknell99, young02}). The extension of this estimation from FR IIs to sources in gas-rich environments, primarily FR Is, became feasible after the launch of the Chandra satellite. \cite{bir04} presented an analysis of 16 galaxy clusters to estimate radio jet powers by assessing the jet to inflate $pV$ energy in cavities. They adopted a buoyancy timescale to estimate cavity ages and found a strong correlation between cavity powers and 1.4 GHz radio luminosities. \citet{cav10} expanded the sample (mainly FR Is) in \citet{bir04} based on the Chandra X-ray and VLA radio data, deriving a mean scaling relation between the kinetic jet power and the radio luminosity. The scaling relation is in line with the model proposed by \citet{wil99}. Later, \citet{hec14} recast these kinetic jet power methods, presenting a best-fitting linear relation between the kinetic jet power and the large-scale NVSS 1.4 GHz radio emissions with the sample of \citet{cav10},
\begin{equation}
    L_{\rm{kin}} = 7 \times 10^{43} f_{\rm{cav}} (L_{\rm{1.4}}/10^{25} \text{W Hz}^{-1}) ^{0.68} {\rm erg/s}, \label{equ1}
\end{equation}
The $L_{\rm{1.4}}$ is the NVSS 1.4 GHz radio luminosity. The $f_{\rm{cav}}$ includes the uncertainties on the physical state of lobes and is suggested to be $f_{\rm{cav}}=4$ from the best linear relation of the data \citep{hec14}. There are evidences that FR 0s also show the presence of compact jets based on LOFAR (LOw-Frequency ARray), VLA, and VLBI observations \citep{che18, bal19, cap20aa642,che21,gio23}, and their multi-band nuclear luminosities are comparable with those of low-power FR Is: these results support that $L_{\rm{kin}}$-$L_{\rm{1.4}}$  relation is also valid for low-power RGs consistent with a FR 0 classification \citep{hec14}.
Here, we followed the line of interpretation of \citet{bal23} and \citet{hec14} for FR 0s and low-power RGs, and computed the kinetic jet powers for our sample by assuming $f_{\rm{cav}}=4$ as a broad balance between AGN heating and radiative cooling in massive clusters for all LERGs. Based on the data scatter of the relation, we assumed an uncertainty on the $L_{\rm{kin}}$ estimate of $\sim$0.7 dex \citep{raf06,bir08,cav10}.

The radio luminosity is obtained by the corresponding K-corrected radio flux [$S_{\nu} (1+z)^{\alpha-1}$] and the luminosity distance ($d_{\rm{L}}$): 
$\nu L_{\nu}=4\pi d_{\rm{L}}^2\nu S_\nu (1+z)^{\alpha-1}$, in which $\alpha$ is the radio spectral index ($S_\nu \sim \nu^{-\alpha}$) and the luminosity distance $d_{\rm{L}}$ is determined by the redshift ($z$):  
\begin{equation}
d_{\mathrm{L}}=(1+z)\frac{ c}{H_{0}} \int_{1}^{1+z} \frac{1}{\sqrt{\Omega_{\mathrm{M}} x^{3}+1-\Omega_ {\mathrm{M}}}} dx, \label{dis}
\end{equation} 
The radio spectral index is assumed to be 0.7 \citep{jac01} in an optically-thin synchrotron regime.

\subsection{Supermassive black hole}

In this paper, the BH mass for nearby LERGs is estimated using the $M_{\rm{BH}}-\sigma_\ast$ relation \citep{tre02}. The stellar velocity dispersions ($\sigma_\ast$) (see \citealt{cap17a, cap17b, bal18}) are obtained from the MPA-JHU SDSS DR7 release of optical spectrum measurements \citep{aba19}. The BH mass is calculated using the formula $\log(M/M_{\odot}) = (8.13 \pm 0.06) + (4.02 \pm 0.32) \log(\sigma_\ast/\sigma_0)$, where $\sigma_0 = 200$ km/s \citep{tre02}. This  $M-\sigma_\ast$ relation has an intrinsic spread in $\log M_{\rm{BH}}$ (BH mass in the unit of solar mass) that is no more than 0.25 to 0.3 dex \citep{tre02}. The three FRCAT samples have comparable BH masses ($\gtrsim$10$^{7.5}$ M$_{\odot}$)\footnote{Apart from three FR II LERGs with M$_{\rm BH}<$10$^{7.5}$ M$_{\odot}$ which could represent a non-negligible radio-quiet AGN contamination in the sample \citep{chi11}.}.

The BH masses and kinetic jet powers are presented in Table \ref{tab:1}, and are both subject to uncertainties arising from the data scatter of their defining relations (0.25 and 0.7 dex, respectively, e.g. \citealt{tre02, cav10}). Here we have used these uncertainties to broadly evaluate the consistency of the relations studied in Fig.~\ref{fig:bp-bz} and Fig.~\ref{fig:jet-edd} with the observed data points.

\section{Jet formation mechanism}
The central jet formation of AGNs is related to the spin of a black hole (BZ model) \citep{bla77} or related to the accretion disc (BP model) \citep{bla82}. Some authors also suggested that a hybrid (BP-BZ) jet model could be adopted for the explanation of the jet formations \citep{mei01,nem07}. Here, we calculated the kinetic jet power, and discussed the possible jet model mechanism for the 431 LERGs based on the self-similar solution of ADAF around a Kerr BH.

(1) The Blandford-Znajek jet model:

According to the BZ model, the jet power is calculated in the consideration of the following formula \citep{nem07, wu08}: 
\begin{equation}
    P^{\rm{BZ}}_{\rm{kin}}=\frac{1}{32} \omega^2_{\rm{F}} B_{\perp}^2 R_{\rm{H}}^2 j^2 c, \label{bz}
\end{equation} 
The $\omega_{\rm{F}}$ is a factor, determined by the relation between the angular velocity of the flux ($\Omega_{\rm{F}}$) threading the horizon and the angular velocity of the black holes ($\Omega_{\rm{H}}$): $\omega_{\rm{F}}=\Omega_{\rm{F}}(\Omega_{\rm{H}}-\Omega_{\rm{F}})/\Omega_{\rm{H}}^2$. The $\omega_{\rm{F}}=1/2$ is for the maximum power outputs \citep{mac82}. The $c$ is the speed of light. The $B_{\perp}$ represents the strength of the magnetic field perpendicular to the horizon, and is assumed to be approximated to the poloidal component of the magnetic field \citep{liv99, nem07}: $B_{\perp}  \approx B_{\rm{p}}(R_{\rm{ms}}) \approx g(R_{\rm{ms}}) B (R_{\rm{ms}})$,  
and $g$ is the field-enhancing shear expressed as $g=\Omega/\Omega'$. 
The $R_{\rm{ms}}$ is the radius at the marginally stable orbit of the accretion disc, which is related to the BH mass and spin (Appendix Eq. \ref{rms}).
The observer at infinity will see an angular velocity $\Omega=\Omega'+\omega$ with the Boyer-Lindquist coordinate system \citep{bar72}, where $\omega$ is the local space-time rotation enforced by the Kerr BH under the same system, determined as shown in Appendix \ref{appA}. $R_{\rm{H}}=(1+\sqrt{1-j^2}) GM_{\rm{BH}}/c^2$ is the horizon radius \citep{nem07}, where $G$ is the gravitational constant, and $j$ is the spin of the black hole.

(2) The hybrid jet model:

The numerical simulations of both the accretion flow and magnetic fields are the essential elements for jet productions \citep{mei01}. Therefore, the jet formation mechanism is possible to be formed by a mix of BP and BZ mechanisms.
In the case of the hybrid (BP-BZ) jet model, the jet power is estimated by \citep{mei01,nem07}:
\begin{equation}
    P^{\rm{Hybrid}}_{\rm{kin}}=\frac{1}{32c}  (B_{\phi} H R \Omega)^2 ,
\end{equation}
in which the azimuthal component of the magnetic field is $B_{\phi} = g(R_{\rm{ms}}) B (R_{\rm{ms}})$, and the vertical half-thickness of the disc is similar to the radii in the self-similar ADAF model \citep{nem07}, $H \approx R$.  The relevant parameters for the above model with the self-similar solution of ADAF are shown in Appendix \ref{appA}. The radius is evaluated at the marginally stable orbit of the accretion disc $R=R_{\rm{ms}}$.

\section{Discussion}
\subsection{Jet power}
\begin{figure}
    \centering
    \includegraphics[width=0.9\linewidth]{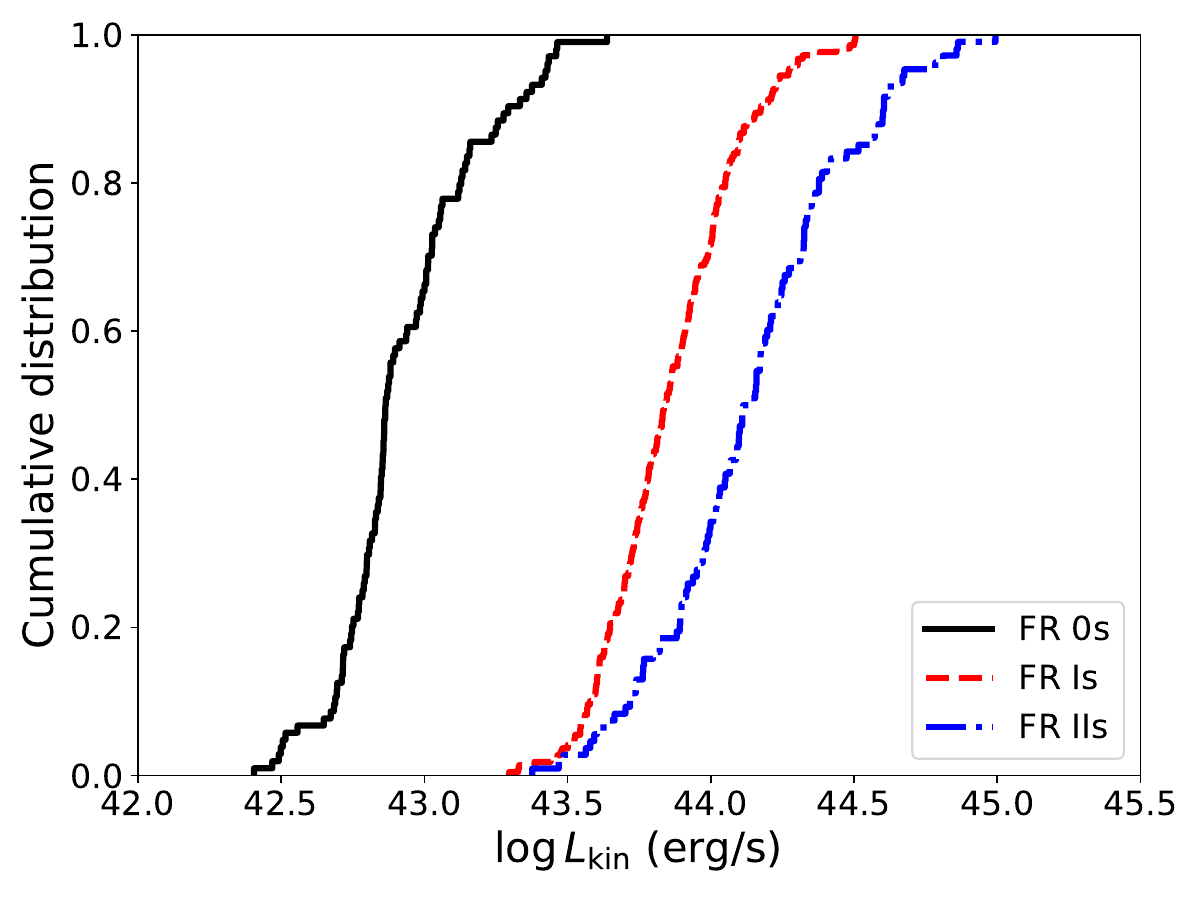}
    \caption{The cumulative distributions of the kinetic jet power for 431 low-excitation radio galaxies. The labels are the same as Fig. \ref{fig:z}. The kinetic jet power is derived from the NVSS 1.4 GHz kiloparsec-scale radio emissions. The average kinetic jet power for FR 0s is $\langle \log L_{\rm{0,kin}}\rangle=42.93\pm0.02$ erg/s, while for FR Is, it is $\langle \log L_{\rm{I,kin}}\rangle=43.86\pm0.02$ erg/s, and for FR IIs, it is $\langle \log L_{\rm{II,kin}}\rangle=44.14\pm0.03$ erg/s.}
    \label{fig:jet}
\end{figure}

In the scenario of ADAF, most of the total jet power is converted into the kinetic power with the minimal radiative power \citep{nem12,hec14}. In the ADAF-dominated galaxies, the total radio luminosity is used as a proxy for estimating the kinetic jet power. We estimated the kinetic jet power within the NVSS 1.4 GHz flux density by following the computations of \citet{hec14} [Eq. (\ref{equ1})]. The cumulative distributions of jet powers for the whole 431 LERGs is shown in Fig. \ref{fig:jet}. 
The average kinetic jet power for FR 0s is $\langle \log L_{\rm{0,kin}}\rangle=42.93\pm0.02$ erg/s, while for FR Is, it is $\langle \log L_{\rm{I,kin}} \rangle=43.86\pm0.02$ erg/s, and for FR IIs, it is $ \langle \log L_{\rm{II,kin}} \rangle=44.14\pm0.03$ erg/s and they are all statistically different.

\subsection{Accretion rate}
The division between FR Is and FR IIs is explained by the differences in host properties or accretion processes. \citet{led96} analysed a statistically complete sample of 188 RGs, and demonstrated that FR Is and FR IIs are distinctly separated by a dividing line (Ledlow-Owen line) between the host mass and the total radio luminosity. \citet{ghi01} suggested that variations in central engines and accretion rates could explain this FR I-FR II dichotomy. Later, \citet{wu08} used the hybrid jet model developed by \citet{mei01}, based on ADAFs around Kerr BHs, to calculate the maximum jet power. They found that the relationship between the jet power and BH mass predicted by the hybrid jet model (e.g. the red dashed line of the hybrid jet model in Fig. \ref{fig:bp-bz}) is distributed similarly to the Ledlow-Owen line, assuming a standard conversion from host mass to BH mass \citep{mcl01,ghi01}.

To evaluate the disc properties,
\citet{wangJM02} defined a line accretion rate between the total line luminosity and Eddington luminosity: 
\begin{equation}
    \lambda = L_{\rm{line}}/L_{\rm{Edd}}, \label{lineacc}
\end{equation}
where $L_{\rm{Edd}}$ is the Eddington luminosity [$L_{\rm{Edd}}=1.38\times 10^{38}(M/M_{\odot})$ erg/s]. The total line luminosity is determined by the fractional contribution of the observed line (e.g. Lyman-alpha line) to the overall line emission \citep{cel97}, $\langle L_{\rm{line}}\rangle = 556 \langle L_{\rm{Ly\alpha}}\rangle$. The fractional line emission ratio $\langle L_{\rm{Ly\alpha}}\rangle:\langle L_{\rm{[O~III]}}\rangle = 100:3.4$  is presented by \citet{fra91} from a high signal-to-noise composite quasar spectrum. We collected the BH mass and [O~III] ($\lambda$5007\AA) emission for 431 LERGs \citep{cap17a,cap17b,bal18} to compute their line accretion rate ($\lambda$) to study disc properties.

\begin{figure}
    \centering
    \includegraphics[width=0.9\linewidth]{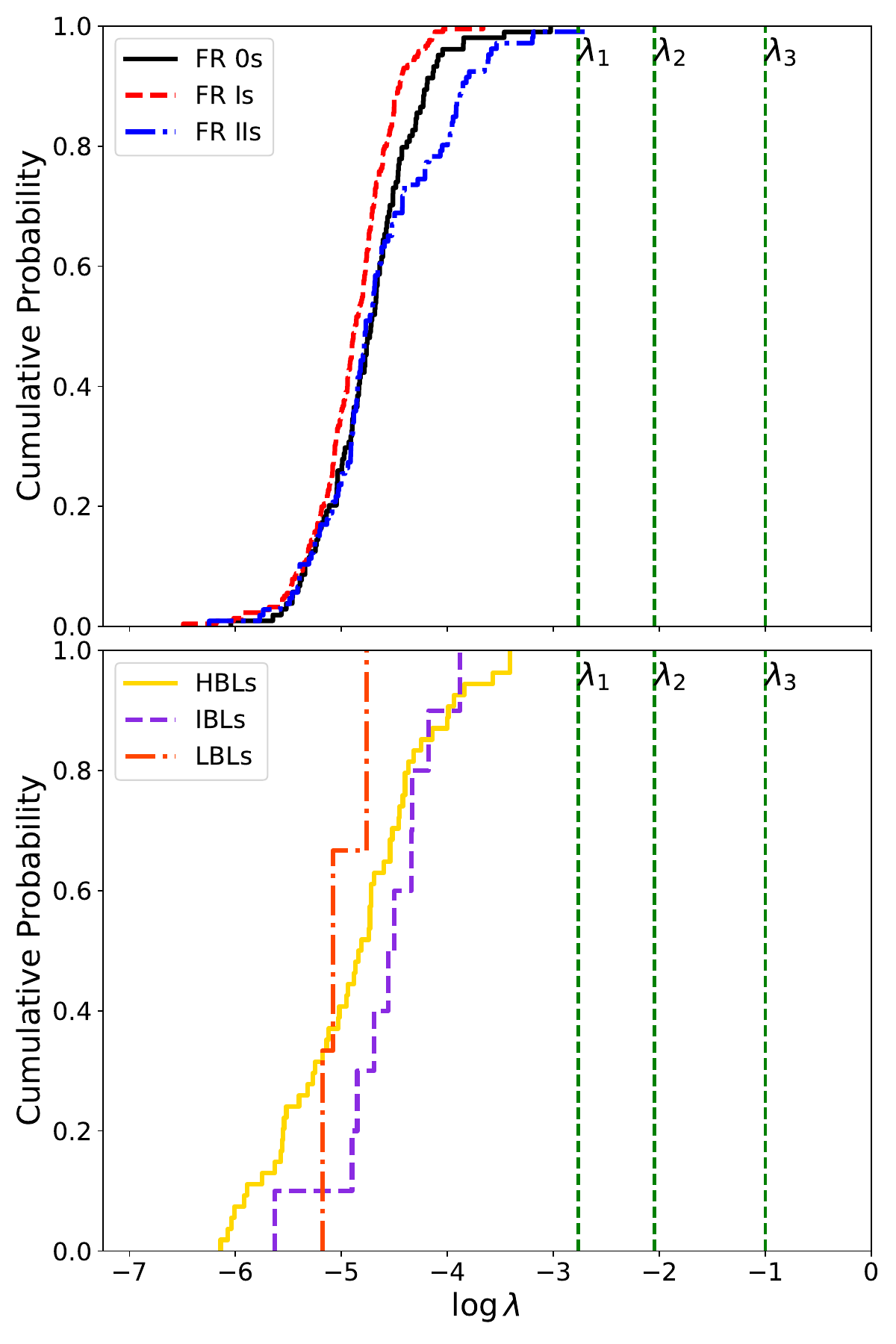}
    \caption{The cumulative  distributions of the line accretion rate [$\lambda=L_{\rm{line}}/L_{\rm{Edd}}$]. The line accretion rate distributions are divided into 4 regions that correspond to different accretion discs: (1) $\lambda<\lambda_1$ is the pure advection-dominated accretion flows (ADAFs); (2) $\lambda_1<\lambda<\lambda_2$ is the coexistence of the ADAFs and the optical thick and geometrically thin disc; (3) $\lambda_2<\lambda<\lambda_3$ is the standard thin disc; (4) $\lambda>\lambda_3$ is the Super-Eddington accretion. The upper panel is for 431 low-excitation radio galaxies. The labels are the same as Fig. \ref{fig:z}. The lower panel is for the 75 Fermi BL Lacertae objects. The yellow solid line is for high synchrotron peaked BL Lacs (HBLs); the purple dashed line is for intermediate synchrotron peaked BL Lacs (IBLs), and the orange dash-dotted line is for low synchrotron peaked BL Lacs (LBLs).}
    \label{fig:lineacc}
\end{figure}

The cumulative distributions of the line accretion rates (in log scale) for the 431 LERGs are presented in the upper panel of Fig. \ref{fig:lineacc}. The average line accretion rate for FR 0s is $\langle \log \lambda_{\rm{0}}\rangle=-4.73\pm0.05$, that for FR Is is $\langle \log \lambda_{\rm{I}}\rangle=-4.89\pm0.03$, and that for FR IIs is $\langle \log \lambda_{\rm{II}}\rangle=-4.64\pm0.06$. Radiatively inefficient discs, such as ADAF type, are characterized by the low accretion rates ($\dot{m} <$ 0.2).

It is assumed that the narrow-line region (where the lines are measured from) is photoionized by the accretion disc\footnote{Although a possible jet contribution in photoionising the medium is not negligible (e.g. \citealt{capetti05}).}, $L_{\rm{line}}=\xi L_{\rm{disc}}$, where $\xi\approx0.1$ \citep{net90,ghi14}. From \citet{mah97}, the total disc luminosity from an ADAF is given by $L_{\rm{disc}} \approx \alpha_{\rm{vis}}^{-2}M_{\rm{BH}}\dot{m}^2$, and the dimensionless accretion rate $\dot{m}=\dot{M}/\dot{M}_{\rm{Edd}}$, $\dot{M}_{\rm{Edd}}=\frac{L_{\rm{Edd}}}{\eta c^2}=1.38 \times 10^{18} M_{\rm{BH}}$(g s$^{-1}$), $\eta=0.1$ represents the accretion efficiency \citep{qua99}. Considering the relation between the disc and total emission line luminosities ($L_{\rm{line}}=\xi L_{\rm{disc}}$), the relation between the line accretion rate ($\lambda$) and the dimensionless accretion rate ($\dot{m}$) for the optically thin ADAFs can be expressed as follows \citep{wangJM02}, 
\begin{equation}
    \dot{m}=2.17\times10^{-2}\alpha_{0.3}\xi_{-1}^{-1/2}\lambda_{-4}^{1/2},
\end{equation}
where $\alpha_{0.3}=\alpha_{\rm{vis}}/0.3$, $\alpha_{\rm{vis}}$ is the viscosity parameter; $\lambda_{-4}=\lambda/10^{-4}$; $\xi_{-1}=\xi/0.1$.  
 
The ADAFs are thought to follow $\dot{m} \leq \alpha_{\rm{vis}}^2$ \citep{qua99}. Therefore, $\lambda_1$ could be rewritten as:
\begin{equation}
    \lambda_1=1.72 \times 10^{-3} \xi_{-1} \alpha_{0.3}^2, 
\end{equation}
Optically thin ADAFs require $\lambda <\lambda_1$. The optically thick, geometrically thin disc (SS) obeys \citep{wang03},
\begin{equation}
    \dot{m}=10 \xi_{-1}^{-1}\lambda,
\end{equation}
and the condition $1>\dot{m} \geq \alpha^2$ gives, 
\begin{equation}
    \lambda_2=9.0 \times 10^{-3} \xi_{-1} \alpha_{0.3}^2,
\end{equation}
A standard thin disc satisfies $\lambda \geq \lambda_2$ \citep{wang03}. When the dimensionless accretion rate reaches $\dot{m}\geq 1$, then we have,
\begin{equation}
    \lambda_3=0.1 \xi_{-1},
\end{equation}

A slim disc requires $\lambda \geq \lambda_3$ \citep{wang03}, namely Super-Eddington accretion.  The region between \(\lambda_1\) and \(\lambda_2\) is considered a transition zone where both ADAF and thin disc can coexist \citep{qua99, ho00}.

The corresponding boundaries are strongly influenced by the viscosity parameter $\alpha_{\rm{vis}}$. Advection models have been successful in explaining BH candidates with $\alpha_{\rm{vis}}$ values in the range $0.1-0.3$ \citep{nar95nature}; the ADAF model for X-ray binaries with BH candidates requires $\alpha_{\rm{vis}} \approx 0.25$ to match observations \citep{qua99}. Following \citet{nar95, mah97, wangJM02, nem07}, we adopted $\alpha_{\rm{vis}} = 0.3$ to analyse the accretion scenarios.
The boundaries for $\lambda$ are marked by the green dashed lines in Fig. \ref{fig:lineacc}. All LERGs but one (SDSS J122640.22+253855.5, $z = 0.134$, falls between the $\lambda_1$ and $\lambda_2$ boundaries) lie below the $\lambda_1$ boundary, suggesting that pure ADAFs dominate their central regions. The K-S tests reveal differences in accretion distributions among FR 0s, FR Is, and FR IIs, with each pair with $p<0.09$.

\subsection{Accretion-ejection paradigm}
Considering the distributions of line accretion rates, most LERGs fall into the pure ADAF region where $\dot{m}<\lambda_1$. Based on it, we explored the BZ and BP-BZ hybrid jet mechanisms for LERGs within the framework of ADAFs around the Kerr BH.  The BP model is generally associated with an AGN with the bright disc \citep{rawlings91}; however, the only BP model is not considered because there is no evidence of luminous disc emissions in the LERG population in general \citep{har09,baldi10,bes12}.

We adopted $\alpha_{\rm{vis}} = 0.3$ for the hybrid jet power calculation, as this value aligns with magnetohydrodynamic simulations in the Kerr metric \citep{wu08}. In fact, two different values of the viscosity parameter, $\alpha_{\rm{vis}}=0.1$ and 0.3, would have a negligible effect on the jet powers estimated from both the BZ and hybrid jet models ($\Delta \log L_{\rm{kin}}\sim 0.1$ erg/s), for a given BH mass ($\log M_{\rm{BH}}=8$), and accretion rate ($\dot{m}=0.01$) \citep{wu08}.  A maximal BH spin of 0.998 \citep{tho74}, and $\dot{m} = 0.01$ were adopted to compute the maximum jet powers (see Appendix \ref{appA}). Additionally,  in the ADAF-dominated sources, the estimated magnetic field strength ($B_{\rm{p}}\approx B_{\perp}$) near the BH is of the order of kGauss for BH masses ($\log M_{\rm{BH}}$) between 8 and 9 (see Appendix \ref{magnetic_field}), consistent with estimated values for FR Is \citep{bla19,kin22,kin24}.

The jet powers are expected to depend on both the spin of the BH, the magnetic field and the structure of the accretion disc. If the radio jet is powered by a rotating BH accreting from a magnetized plasma, this mechanism extracts rotational energy from the magnetic field lines threading the spinning BH,  in which the relation between the BZ jet power and the strength of the magnetic field perpendicular to the horizon ($B_{\perp}$) satisfies that $P^{\rm{BZ}}_{\rm{kin}}\propto B_{\perp}^2 M^2_{\rm{BH}}$  \citep{fra92}. This implies that more massive and highly magnetized BHs can generate stronger jets, with the efficiency mainly depending on the spin of the BH. On the other hand, when the radio jet power is contributed by both a rotating accretion disc and BH via magnetic fields anchored inside and outside the ergosphere, 
the jet power is related to the azimuthal component of the magnetic field ($B_{\phi}$) and BH mass, following the relation: $P^{\rm{Hybrid}}_{\rm{kin}}\propto B_{\phi}^4 M^2_{\rm{BH}}$, in which $H\sim R$ for ADAF-dominated sources.

\begin{figure}
    \centering
    \vspace{-1cm}
    \includegraphics[width=1\linewidth]{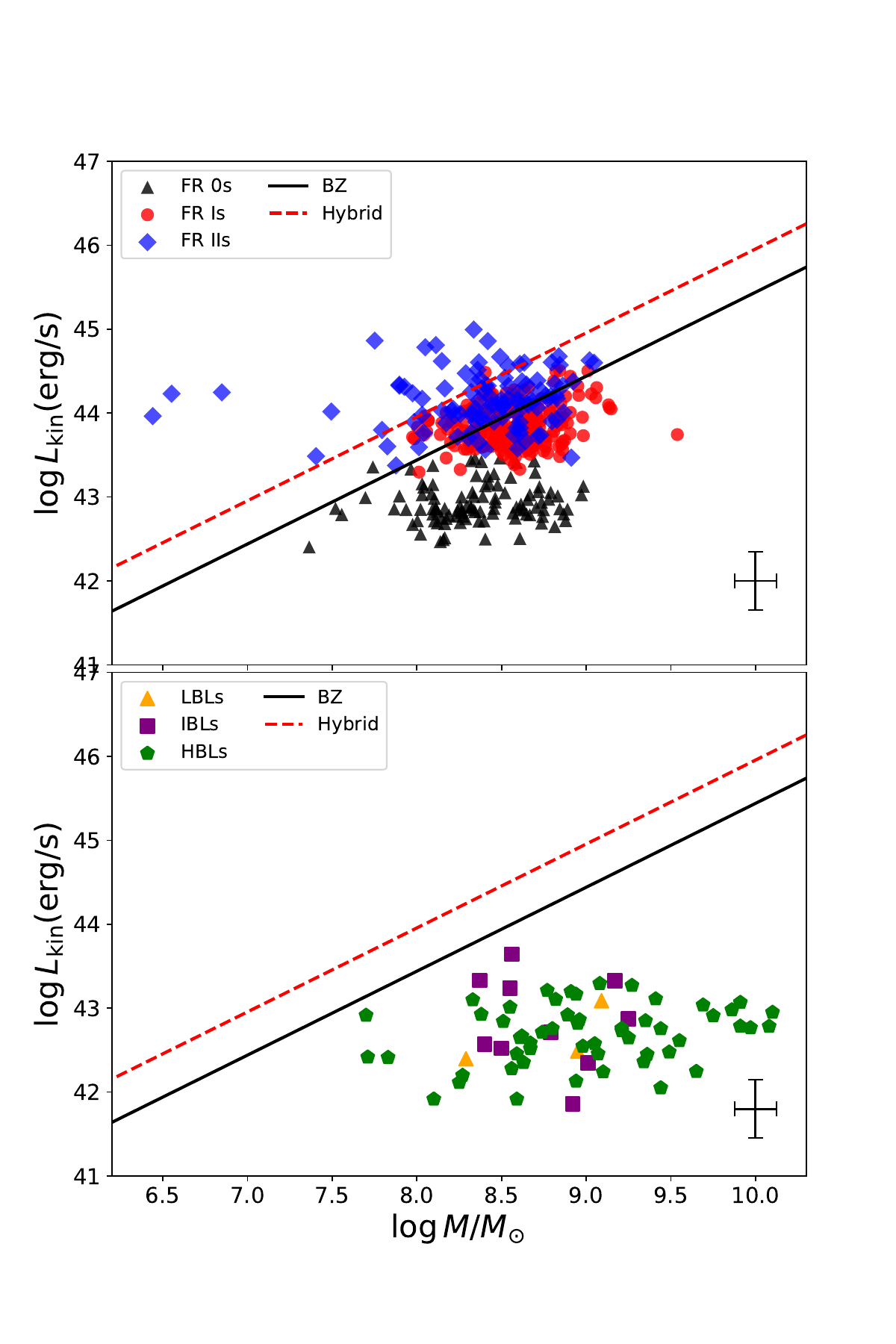}
    \vspace{-1cm}
    \caption{The kinetic jet power versus black holes for (1) the upper panel: 431 low-excitation radio galaxies. the black triangle is for FR 0s, the red circle is for FR Is, and the blue diamond is for FR IIs. (2) the lower panel: 75 Fermi BL Lacertae objects. The yellow triangle is for LBLs, the purple square is for IBLs, and the green pentagon is for HBLs. Their intrinsic kinetic jet power is obtained assuming a Doppler factor $\delta\sim10$ within the continuous jet model ($p=2+\alpha$) and the flat spectral index ($\alpha=0$). The black solid line is for the maximum jet power in the Blandford-Znajek (BZ) jet model; the red dashed line is the mixture (Hybrid) of the BZ and Blandford-Payne jet model; 
    The average uncertainty for BH mass is 0.25 dex, and the average jet power uncertainty is 0.7 dex.}
    \label{fig:bp-bz}
\end{figure}

The relationship between the jet power and BH mass is presented in the upper panel of Fig. \ref{fig:bp-bz}.  The black solid line is for the maximum output of BZ jet model, while the red dashed line is for the maximum output of BP-BZ hybrid model. Around 99\% (103/104) of FR 0s lie below the pure BZ line, suggesting that the jet powers of FR 0s are most likely explained by the BZ mechanism, although an explanation from the BP-BZ hybrid mechanism cannot be entirely excluded (e.g. spine-sheath jet structure; see \citealt{bal23} and references therein). Notably, we disfavour BP mechanism for FR 0s because there are no signatures of bright standard thin disc in FR 0s based on the X-ray properties \citep{tor18}. Therefore, conservatively, the BZ-based jet coupled with an ADAF-type disc is more in agreement with the general radio and X-ray characteristics of FR 0s.

Compelling multi-band observational evidence points to the presence of an ADAF disc at the centre of FR Is (e.g. \citealt{chiaberge99,bal06,hardcastle07,he24}). \citet{cao04} proposed that the BZ mechanism alone does not provide enough power to explain the high radio jet power for at least a third of the FR Is in their sample.
In this work, we found that $\sim99\%$ FR Is (218/219) fall below the BP-BZ hybrid line, with 146 of these 218 FR Is also falling below the BZ line. This suggests that the BZ mechanism by itself cannot explain the all FR I sample, and that a mix of pure BZ and BP-BZ jet mechanisms and/or higher magnetic intensity could reduce the tension with the results, as seen by \citet{cao04}.

However, only 23 of 108 FR IIs fall below the BZ jet model line, indicating that the jet power of majority FR IIs cannot be explained by the pure BZ model; almost half sample of FR IIs (58/108) located between the BZ and hybrid line; their accretion disc may contribute to support their jets and explain their kinetic jet power. Additionally, 27 out of 108 FR IIs lie above the BP-BZ hybrid jet power line. The BP-BZ hybrid jet mechanism with ADAF scenario are not able to explain the higher jet power for, at least, a third of the FR II sample. The higher jet power may be explained by the magnetization-driven outflows from the accretion disc \citep{cao13}. \citet{cao18} compared the jet powers of blazars with the BZ and BP jet mechanisms, and presented that the maximal jet powers from both BZ and BP models are always not sufficient to explain the strong jets in some blazars. Therefore, the author proposed that large magnetic fields dragged inwards by the accretion disc with magnetized outflows may support and accelerate the jets and explain their high kinetic powers. Analogously, this mechanism may account for the high jet luminosities observed in some BL Lacs \citep{chen23mn}, as well as in this subset of FR IIs.

Even though the higher jet power in some FR Is and FR IIs could be explained by the hybrid BP-BZ mechanisms and/or the magnetization-driven outflows, however, the observed spread of the jet powers among FR 0s, FR Is and FR IIs could also be explained by the differences in either BH spins or magnetic field intensities under the framework of the BZ jet mechanism \citep{gra21}. Both the BZ and the hybrid BP-BZ jet power are proportional to the BH mass, BH spin, and magnetic field, but the dynamic range of BH spin values (0-1) is much smaller than the observed range of magnetic fields involved in the jet (some order of magnitudes\footnote{A wide range of magnetic field intensity are observed in RGs ranging (as $B\sim r^{-1}$) between the central engine to the jet, from kGauss to tens $\mu$Gauss (e.g. \citealt{dalla02, cro05, sta05, sta_kat05, bru13, ori20,kin22,kin24}).}). Therefore, since BH masses among the various FR classes are similar, it is natural to expect that the magnetic field might be the main ingredient in supporting the most powerful jets \citep{nem07}. Unfortunately, accurately estimating the magnetic field is challenging, and how the magnetic fields evolve in the jet base among FR 0s, FR Is, and FR IIs is unclear. Figure \ref{fig:bp-bz} shows the maximum BZ (solid black line) jet power for a jet with maximized BH spin ($j=0.998$) by considering the magnetic field at the marginally stable orbit of the accretion disc. If we assume that FR 0s, FR Is and FR II LERGs consist in a single population of RGs with a continuous distribution of properties, which share the same BZ jet mechanism \citep{bal23}, a gradual increase of the magnetic field of the jet base could lead to a gradual increase of the jet power across the FR 0-FR I-FR II LERG population. In fact, the lower magnetic field intensity in FR 0s than FR Is has been invoked to justify the failure to develop a powerful jet for the former \citep{gra21,bal23}, and the stronger magnetic field in FR IIs than FR Is has instead been invoked to account for different SED fitting
 with synchrotron self-Compton model \citep{xue17} or different accretion-ejection efficiencies \citep{gra21}.
 
\subsection{Unification Scenario}
The unified scheme for blazars and RGs has been studied for decades (e.g. \citealt{urr95}).  BL Lacs are suggested to be the beamed populations of FR Is, and FSRQs are the beamed populations of FR IIs \citep{pad92, ghi93, xie93, urr95, chi00}. The accretion-rate separations for blazars are discussed in \citet{ghi11} and \citet{sba12}. BL Lacs exhibit lower jet powers and accretion rates, clearly separating them from FSRQs. An accretion-based unification of blazars and RGs is discussed in the works \citep{xu09, sba14, che15}. If blazars are RGs viewed at a small jet viewing angle, they should share similar accretion discs and jet formation mechanisms.

Thanks to Fermi-LAT (Fermi Large Area Telescope), 3814 AGNs were observed in the latest LAT 12-year Source Catalogue (4FGL-DR3) \citep{aje22}. \cite{yang22} compiled a large sample of 2709 Fermi-LAT blazars with multi-band data and fitted the first peak of the SED with a parabolic equation to discuss their jet properties and acceleration mechanisms. Based on the Bayesian classification of the synchrotron peak frequencies, they classified the BL Lacs into low, intermediate and high synchrotron peaked BL Lacs (analogously to blazar classifications, see Sect.~\ref{intro}). Because some BL Lacs exhibited the broad emission lines, akin to FSRQs \citep{ghi11}, \citet{pal21} divided a Fermi sample of 1077 blazars into emission- and absorption-line blazars based on the optical spectroscopic information. They presented a negative tendency for Fermi blazars between the disc accretion rate ($L_{\rm{disc}}/L_{\rm{Edd}}$) and synchrotron peak frequency ($\nu_{\rm{peak}}$). The majority of emission-line blazars are predominantly low synchrotron peaked sources (e.g. FSRQs and LBLs), and are characterized by broad emission lines and high disc accretion rates,  whereas absorption-line blazars, are predominantly intermediate or high synchrotron peaked sources (e.g. IBLs and HBLs) exhibiting prominent absorption lines and low disc accretion rates \citep{pal21}.

As a consistent comparison to LERGs, we compiled a redshift-limited ($z<0.15$) sample of 75 Fermi absorption-line BL Lacs (61 HBLs, 11 IBLs, and 3 LBLs) with observed 1.4 GHz radio luminosities and synchrotron peak classifications from \citet{yang22}, and disc luminosities\footnote{The disc luminosity for absorption-line BL Lacs is obtained from the $3\sigma$ upper limit of the total emission-line luminosity.} and BH mass estimations (from the stellar velocity dispersion\footnote{$\log(M/M_{\odot}) = (8.12 \pm 0.08) + (4.24 \pm 0.41)\times \log(\sigma_\ast/200 \text{ km s}^{-1})$, \citet{gul09}.}) from \citet{pal21}. Given the Doppler beaming effect in radio jets, the intrinsic radio luminosity for BL Lacs follows, $\log L_{\rm{in}} = \log L_{\rm{ob}} - p\log \delta$, where $p = 2 + \alpha$ is for a continuous jet model, and  $p = 3 + \alpha$ is for a spherical jet model \citep{urr95}. The kinetic jet power for BL Lacs was calculated using Eq. (\ref{equ1}), with the intrinsic radio luminosity based on an average Doppler factor of $\delta \approx 10$ \citep{lio18,hom21} in the case of a continuous jet model ($p = 2 + \alpha$) with a flat spectral index ($\alpha = 0$) (see e.g. \citealt{cap02}). All the parameters are presented in Table \ref{tab:1}.    

\begin{figure}
    \centering
    \includegraphics[width=0.9\linewidth]{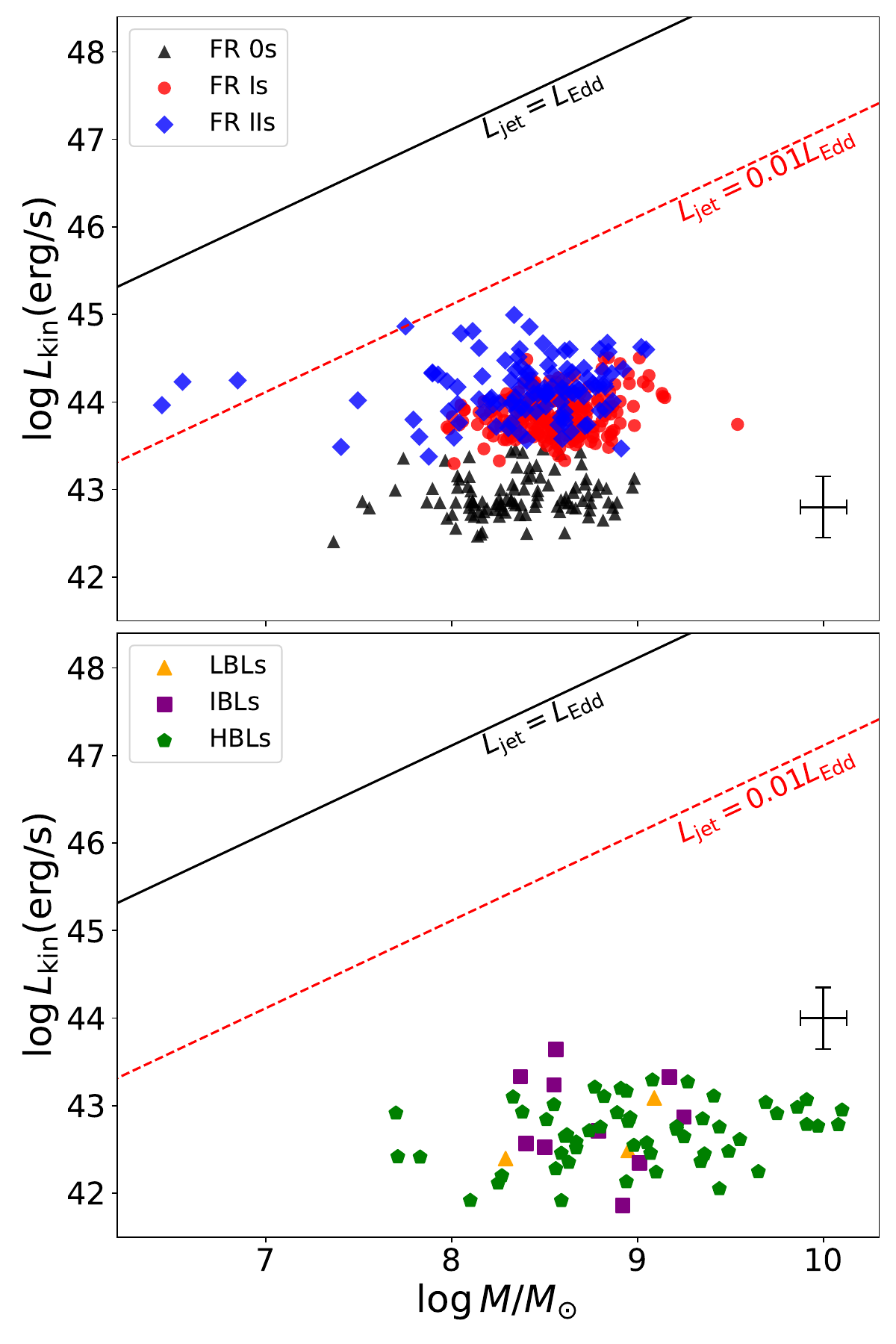}
    \caption{The plot of supermassive black hole versus the kinetic jet power with the lines of the $L_{\rm{kin}}=L_{\rm{Edd}}$ (black solid line) and $L_{\rm{kin}}=0.01 L_{\rm{Edd}}$ (red dotted line). (1) the upper panel is for 431 low-excitation radio galaxies; (2) the lower panel is for 75 Fermi BL Lacertae objects. The labels are the same as the Fig. \ref{fig:bp-bz}. The average uncertainty for BH mass is 0.25 dex, and the average jet power uncertainty is 0.7 dex.}
    \label{fig:jet-edd}
\end{figure}

Figure \ref{fig:jet-edd} depicts the kinetic jet powers for our LERG sample and 75 Fermi BL Lacs, which all fall below the line of $L_{\rm{kin}}=0.01 L_{\rm{Edd}}$, in agreement with the results of \citet{ghi01} and \citet{xu09}. LERGs and BL Lacs are expected to have both low-accretion sources. Therefore, we computed the line accretion rate ($\lambda$) for BL Lacs [Eq. (\ref{lineacc})]. The cumulative distributions of $\lambda$ for our BL Lac sample are shown in the lower panel of Fig. \ref{fig:lineacc}. The line accretion rates for 75 BL Lacs are smaller than the boundary of $\lambda_1$, suggesting that the central discs for the absorption-line BL Lacs (mostly HBLs) are dominated by ADAFs.

Our results agree with previous studies on BL Lacs: i) \citet{chen23mn}, studying the accretion properties of the BL Lac sample from \citet{pal21}, found that IBLs and HBLs are in the optically thin ADAF state. ii) \citet{zhao24}, studying the energy budget in the jets of a sample of 348 Fermi HBLs, concluded that HBLs may have optically thin ADAF in their centres due to their very low radiation efficiency. A larger and more complete sample of IBLs and LBLs is needed to compare robustly the accretion properties with LERGs.

When discussing the relationship between kinetic jet powers and BH mass, and considering the jet mechanism for BL Lacs, the intrinsic jet power for 75 BL Lacs (mostly HBLs) can be explained by the BZ mechanisms (Fig. \ref{fig:bp-bz}). Based on our results, BZ model appears sufficient to justify the kinetic jet power for the absorption-line BL Lacs (mostly HBLs). However, due to the limited number of LBLs and IBLs in the sample, a solid conclusion cannot be drawn for these subclasses.

In summary, our and previous results indicate a possible affinity between HBLs and low-$\dot{m}$ RGs.

\subsubsection{FR 0 LERGs and BL Lacs}

On one hand, compact radio sources hosted in red massive early-type galaxies have been named as FR 0 RGs \citep{sadler14,bal16}. The nuclei of these FR 0s show evidence of ADAF-dominated disc, and are classified as LERGs \citep{tor18,baldi19rev,bal23}. On the other hand, LERGs and HERGs are  proposed to be the parent populations for BL Lacs and FSRQs, respectively \citep{lai94,gio12}. As a subclass of LERGs, FR 0s are expected to align with the unified scheme for BL Lacs.

In line with this scenario, \citet{mas20},  studying the cluster richness difference between BL Lacs and RGs, noted that the large-scale environmental properties of 14 nearby BL Lacs are statistically consistent with that of FR 0s, suggesting that FR 0s could be the parent population of the BL Lacs. In addition, both FR 0s and BL Lacs share similar intrinsically-low jet powers, weaker than those of their relative RG 'cousins', FR I-IIs and FSRQs.

Furthermore, we cross-checked the 14 nearby BL Lacs of \cite{mas20} with the synchrotron classification of \citet{yang22}, and 11 of 14 are matched within the name and the redshift. The synchrotron classification and the synchrotron peak value for the 11 BL Lacs are listed in (4) and (5) of Table \ref{tab:2}. 10/11 BL Lacs are classified as HBLs with the frequency range of $\log \nu_{\rm{peak}}=15.5-17.3$ Hz. Meanwhile, one BL Lac (5BZBJ1203+6031) falls into the classification of IBL with a synchrotron peak of $\log\nu_{\rm{peak}}=14.8$ Hz, which is close to the high synchrotron peaked boundary ($\log \nu_{\rm{peak}}=14.9$ Hz, \citealt{yang22}). These results support that the HBLs may reasonably be regarded as the beamed counterparts of nearby FR 0s: both of their centres are dominated by the ADAFs (low-$\dot{m}$ in general), hosted in early-type galaxies, their jet powers can be explained by the BZ mechanism, and they share the similar large-scale environments \citep{mas20}.

\subsubsection{FR I-II LERGs and BL Lacs}
Concerning the FR I LERGs,  the classical unification between FR Is and BL Lacs has been discussed for decades \citep{pad92,xie93,urr95,capetti99,chen15aj}. However, the beamed counterparts of FR Is may not belong to a single subclass of BL Lacs \citep{mer11}. \citet{gir04} found that BL Lacs, particularly HBLs, have intrinsic core radio powers similar to those of FR Is. In a study of a large sample of radio-loud AGNs, \citet{mer11} explored the blazar sequence envelope and its unification. They identified two populations on the $\nu_{\rm{peak}}-L_{\rm{peak}}$ plane: a strong-jet population, consisting of FSRQs and high-power LBLs, which will evolve into FR IIs as the jet viewing angle increases; and a weak-jet population, including IBLs, HBLs, and low-power LBLs, which are expected to evolve into the FR I population as the jet viewing angle increases [e.g. the Fig. (4) and Table (4) of \citealt{mer11}]. In this unification scheme, a stratified jet structure (fast jet spine and a slower layer, \citealt{ghisellini05,bou23}) common between FR Is and BL Lacs would explain their different observed multi-band properties \citep{chi00,capetti02}: a mix of HBLs and LBLs could be the parental population of FR I LERGs \citep{bai02}.

For FR II LERGs, unfortunately, the attempts for a unification with BL Lacs still wanders in an uncharted territory because of the limitation of the data,  sample completeness and homogeneity (or clustering properties, e.g. \citealt{pesce93,falomo93,muriel16}). Some studies showed that the LBLs with intrinsic jet power ($\log L_{\rm{kin}}>44.6$ erg/s), comparable to those of FSRQs, and weak lines could be the beamed counterpart of the FR II LERG population \citep{gio12,fan19}. These LBLs (assuming the BH mass, $\log M_{\rm{BH}}$, ranges from 8 to 9.5) cannot be completely explained by the hybrid jet model with the minimum required magnetic field (e.g. red dashed line in Fig. \ref{fig:bp-bz}), but a magnetization-driven outflow may account for their intrinsic jet power, as observed in some BL Lacs with high jet power \citep{cao18,chen23mn}, which is similar to our study of the one-third FR II LERG sub-sample.

The common characteristics which may point to a kinship between the LERG and BL Lac populations in general can be summarized below: similar low-accretion mode (ADAF type) and lack of evidence of luminous standard accretion disc (e.g. no broad emission lines), low intrinsic jet power explained by a BZ model, and generally rich  Mpc-scale environment (e.g. \citealt{gen13,mas19,mas20}). In line with this commonality, \citet{mooney21} found that
the size and luminosity  distribution of the extended radio emission at 144 MHz of BL Lacs are consistent with expectations based on the AGN unification paradigm, as they are the aligned
counterparts of LERGs. However, a more statistically robust multi-band analysis on a flux-limited samples of FR I-FR II LERGs and BL Lacs is needed to draw solid conclusion on their affinity.  This tentative unification, however, does not consider the life cycle of RG populations (e.g. FR II LERGs might evolve from HERGs, \citealt{tad16,mac20}), which complicates further this scenario.

This proposed unifying scenario would reconcile with the cooling model as an explanation of the blazar sequence, in which the blazars with the stronger (intrinsic) jet power would display the lower synchrotron peak spectrum because of the strong cooling effect \citep{ghi98, pra22}.  While for the higher (intrinsic) jet power in FR IIs, the stronger cooling effect will make the jet particles lose more energy and make their synchrotron peak frequencies of the beamed counterparts of FR IIs fall in the lower frequency part, which, ideally, is expected to be more populated by LBLs \citep{capetti02}.

\begin{table}
	\centering
	\caption{The physical parameters for 14 nearby ($z<0.15$) BL Lacertae objects from Massaro et al. (2020). }
	\label{tab:2}
	\begin{tabular}{lcccc} 
 \hline
 
Name&4FGL name&$z$&Class& $\log \nu_{\rm{peak}}$ \\
 &&&&Hz\\
  (1)&(2)&(3)&(4)&(5)\\
		\hline
5BZBJ0809+5218	&	0809+5218	&	0.138	&	HBL	&	16.2	\\
5BZBJ1058+5628	&	1058+5628	&	0.143	&	HBL	&	15.5	\\
5BZBJ1104+3812&1104.4+3812&0.03&HBL&16.6\\
5BZBJ1117+2014	&	1117+2013	&	0.138	&	HBL	&	15.9	\\
5BZBJ1136+6737	&	1136+6736	&	0.136	&	HBL	&	16.9	\\
5BZBJ1203+6031	&	1203.1+6031	&	0.065	&	IBL	&	14.8	\\
5BZBJ1221+2813	&	1221.5+2814	&	0.102	&	HBL	&	15.5	\\
5BZBJ1257+2412	&			            &	0.141	&	   &			\\ 
5BZBJ1428+4240	&	1428+4240	&	0.129	&	HBL	&	17.3	\\
5BZBJ1534+3715	&	1534.8+3716	&	0.143	&	HBL	&	15.5	\\
J160708.14+592	&			            &	0.132	&		  &		\\
J163738.24+300	&			            &	0.0786	&			&		\\
5BZBJ1653+3945	&	1653.8+3945	&	0.033	&	HBL	&	16.8	\\
J222028.72+281	&	2220.5+2813	&	0.148	&	HBL	&	15.6	\\

		\hline
	\end{tabular}
\medskip

\small
Notes: Col. (1) is the source name; Col. (2) the 4FGL name, Col. (3) the redshift; Col.(4) the synchrotron frequency-peak classification, HBL for the high synchrotron peaked BL Lacs, and IBL for the intermediate synchrotron peaked BL Lacs; Col. (5) the synchrotron peak value from \cite{yang22}.
\end{table}

\section{Conclusions}
We compiled a sample of 431 LERGs with available 1.4 GHz radio and emission-line luminosities, and BH masses to investigate their accretion discs and jet formation mechanisms. We also compared the accretion properties of 75 Fermi BL Lacs with those of the 431 LERGs. We come to the following conclusions:

\begin{itemize}

\item There is a sequence of kinetic jet powers from FR 0s, to FR Is to FR IIs, with FR 0s having an order of magnitude smaller kinetic jet power than FR Is and FR IIs.

\item Almost all the line accretion rates for LERGs are below the pure ADAF disc boundary ($\lambda<\lambda_1$), supporting the idea that LERGs have ADAF-dominated flows. 

\item Considering ADAFs surrounding Kerr BHs, we computed the maximum jet power for the BZ and the BP-BZ hybrid jet models, and found that the kinetic jet power of FR 0s can be explained by the pure BZ model. The kinetic jet power of FR Is can be explained by both the BZ model and the BP-BZ hybrid jet models, whereas the kinetic jet power for about one third of FR IIs could not be explained by the BP-BZ hybrid model. The possible explanations are that the high jet power for this one-third FR II sub-sample could be explained by a magnetization-driven outflow from the accretion disc, or could be covered by the BZ jet model if we considered a more intense magnetic field, as indicated in \citet{gra21}.

\item Both BL Lacs (predominantly HBLs) and LERGs are below the \(L_{\rm{kin}} = 0.01 L_{\rm{Edd}}\), and their centres are likely dominated by ADAFs.  By cross-checking the BL Lac sample of \citet{mas20} and SED classification \citep{yang22}, 11 out of 14 BL Lacs are confirmed as high synchrotron-peaked BL Lacs. Several arguments support that the high-frequency peaked BL Lacs (i.e. HBLs) could be the beamed counterparts of FR 0s, reinforcing the cooling model as an explanation for the blazar sequence.  

\item We tentatively proposed a possible unification between the LERG population and BL Lacs, based on current partial available results: HBLs, IBLs, high-power LBLs could be the beamed counterparts of the whole LERG population, embracing different radio morphological classes FR 0, FR I, and FR II.

\end{itemize}

On one hand, the LERG population (FR 0s, FR Is, FR IIs) show observational evidence of establishing a single continuous population, with similar BH masses, galaxy types, accretion properties, and environments, regardless of their different jet morphologies \citep{bal23}. On the other hand, BL Lacs show affinity more with the LERG population rather than with HERG, which we have highlighted in this work. However, a comprehensive study with a more complete BL Lac sample would provide more robust results regarding the accretion-based unification between BL Lacs and LERGs.

\section*{Data availability}
Tables \ref{tab:1} is only available in electronic form at the CDS via anonymous ftp to \url{cdsarc.u-strasbg.fr (130.79.128.5)} or via \url{http://cdsweb.u-strasbg.fr/cgi-bin/qcat?J/A+A/}.


\begin{acknowledgements}
we greatly appreciate the comments from the anonymous referee, which have helped us improve the  manuscript. The work is partially supported by the National Natural Science Foundation of China (NSFC 12433004, U2031201, NSFC 11733001), the Eighteenth Regular Meeting Exchange Project of The Scientific and Technological Cooperation Committee between the People’s Republic of China and the Republic of Bulgaria (Series No. 1802). We also acknowledge the science research grants from the China Manned Space Project with NO. CMS-CSST-2021-A06, and the support for Astrophysics Key Subjects of Guangdong Province.  R.D.B. acknowledges financial support from INAF mini-grant \textit{\lq\lq FR0 radio galaxies\rq\rq} (Bando Ricerca Fondamentale INAF 2022). This research has made use of the NASA/IPAC Extragalactic Database (NED) which is operated by the Jet Propulsion Laboratory, California Institute of Technology, under contract with the National Aeronautics and Space Administration. X. H. Ye acknowledges the support (NO.202208440164) from the Chinese Scholarship Council.
\end{acknowledgements}

%
\bibliographystyle{aa}
\bibliography{yebib}


\begin{appendix}
\section{Derivation of the jet power} \label{appA}
The self-similar ADAF structure is described by \cite{nar95}. In this paper, we followed the idea of self-similar solution ADAFs structure from \citet{nar95} to quantify the BZ and hybrid BP-BZ mechanism for both LERGs and BL Lacs. The parameters of self-similar ADAF structure as discussed below: 
\begin{equation}
    \Omega'=7.19\times10^4 c_2 M_{\rm{BH}}^{-1} r^{-3/2} \text{s}^{-1},
\end{equation}
\begin{equation}
    B=6.55\times10^8 \alpha^{-1/2} (1-\beta)^{1/2} c_1^{-1/2} c_3^{1/4} M_{\rm{BH}}^{-1/2} \dot{m}^{1/2} r^{-5/4} \text{G}, \label{magnetic_field}
\end{equation}
\begin{equation}
    H\approx (2.5c_3)^{1/2} R.
\end{equation}
where the $\Omega'$ is the angular velocity of the disc; $M_{\rm{BH}}$ is the BH mass in solar mass unit ($M_{\rm{BH}} = M/M_{\odot}$); $r$ is the radii in Schwarzschild units ($r=R/(2GM/c^2)$), where $c$ is the speed of light, and $G$ is the gravitational constant. $B$ is the magnetic field; $\dot{m}$ is the dimensionless accretion rate; $H$ is the vertical half-thickness of the disc.

The constants for $c_1$, $c_2$, $c_3$ are as \citep{bar72}:
\begin{equation}
    c_1=\frac{5+2\epsilon'}{3\alpha^2} g'(\alpha, \epsilon'),
\end{equation}
\begin{equation}
    c_2=[\frac{2\epsilon(5+2\epsilon')}{9\alpha^2} g'(\alpha, \epsilon') ]^{1/2},
\end{equation}
\begin{equation}
    c_3=c_2^2/\epsilon',
\end{equation}
in which both the $g'(\alpha_{\rm{vis}}, \epsilon')$ and  $\epsilon'$ are the variable replacement expressions:
\begin{equation}
     g'(\alpha_{\rm{vis}}, \epsilon') \equiv [1+\frac{18\alpha_{\rm{vis}}^2}{(5+2\epsilon')^2}]^{1/2} -1,
\end{equation}
\begin{equation}
    \epsilon' \equiv \frac{1}{f} (\frac{5/3-\gamma}{\gamma-1}),
\end{equation}
here $f$ is the advection parameter that represents the fraction of viscously dissipated energy \citep{nar95, nem07}, and $\gamma$ is the ratio of specific heats \citep{nar95}.

The relationships among $\alpha_{\rm{vis}}$, $\beta$, and $\gamma$ are defined as follows, 
\begin{equation}
    \gamma=\frac{5\beta+8}{3(2+\beta)},
\end{equation}
\begin{equation}
    \alpha_{\rm{vis}}\approx0.55/(1+\beta),
\end{equation}
Where $\alpha_{\rm{vis}}$ is the viscosity parameter,  and $\beta$ is the ratio of gas to magnetic pressure.

The angular velocity of the field seen by an outside observer at infinity in the Boyer-Lindquist frame is $\Omega=\Omega'+\omega$: where $\omega$ is determined by \citep{bar72}: 
\begin{equation}
    \omega=\frac{2jM}{j^2(R+2M)+R^3},
\end{equation}
with the geometrized units ($G=c=1$). The $j$ is the spin of the black hole. 

Within the consideration of ADAF-dominated flows, the $H \sim R$, and $B_{\rm{p}}\approx H/R B_{\phi} \approx B_{\phi}$.  The azimuthal component of the magnetic field is determined by $B_{\phi}=g(R)B(R)$.
The all the quantities defined in the equations at the marginally stable orbit of the accretion disc $R=R_{\rm{ms}}$.
\begin{equation}
    R_{\rm{ms}}=GM/c^2  \{3+Z_2-[(3-Z_1)(3+Z_1+2Z_2)]^{1/2}\}, \label{rms}
\end{equation}
\begin{equation}
    Z_1 \equiv 1+(1-j^2)^{1/3}[(1+j)^{1/3}+(1-j)^{1/3}],
\end{equation}
\begin{equation}
    Z_2 \equiv (3j^2+Z_1^2)^{1/2},
\end{equation}
A maximal BH spin of 0.998 \citep{tho74}, $\alpha_{\rm{vis}} = 0.3$ \citep{nem07}, $f = 1$ \citep{nem07}, and $\dot{m} = 0.01$ \citep{wu08,chen23mn} were adopted to compute the maximum jet powers.

\FloatBarrier 
\clearpage

\end{appendix}
\end{document}